\documentclass[conference]{IEEEtran}
\IEEEoverridecommandlockouts
\usepackage{xcolor,colortbl}
\definecolor{Gray}{gray}{0.95}
\RequirePackage{lettrine}

\ifCLASSINFOpdf
  \usepackage[pdftex]{graphicx}

 \DeclareGraphicsExtensions{.pdf,.jpeg,.png}
\else
  \usepackage[dvips]{graphicx}
  \DeclareGraphicsExtensions{.eps}
\fi

\usepackage{amssymb,amsmath}
\usepackage[noadjust]{cite}
\usepackage[linesnumbered,ruled,vlined,boxed,commentsnumbered]{algorithm2e}
\usepackage{array,multirow}
\usepackage{caption}
\usepackage{pifont}
\usepackage{mathtools}
\usepackage{caption}
\usepackage{subcaption}
\usepackage{stfloats}

\DeclareRobustCommand*{\IEEEauthorrefmark}[1]{%
  \raisebox{0pt}[0pt][0pt]{\textsuperscript{\footnotesize #1}}%
}

\begin{document}

\title{On-Demand Routing for Urban VANETs using Cooperating UAVs}

\author{Omar Sami Oubbati\IEEEauthorrefmark{1}, Noureddine Chaib\IEEEauthorrefmark{1}, Abderrahmane Lakas\IEEEauthorrefmark{2}, and Salim Bitam\IEEEauthorrefmark{3}\\
\IEEEauthorblockA{\IEEEauthorrefmark{1} Laboratory of Computer Science and Mathematics, University of Laghouat, Algeria\\ \{s.oubbati, n.chaib\}@lagh-univ.dz}
  \IEEEauthorblockA{\IEEEauthorrefmark{2} College of Information Technology, United Arab Emirates University, PO Box 17551, Al Ain, UAE\\ alakas@uaeu.ac.ae}
    \IEEEauthorblockA{\IEEEauthorrefmark{3} LESIA Laboratory, Department of Computer Science University of Biskra, Algeria\\ s.bitam@univ-biskra.dz}
}
\maketitle

\begin{abstract}
Vehicular ad hoc networks (VANETs) are characterized by frequent routing path failures due to the high mobility caused by the sudden changes of the direction of vehicles. The routing paths between two different vehicles should be established with this challenge in mind. Stability and connectedness are a mandatory condition to ensure a robust and reliable data delivery. The idea behind this work is to exploit a new reactive routing technique to provide regulated and well-connected routing paths. Unmanned Aerial Vehicles (UAVs) or what are referred to as drones can be both involved in the discovery process and be full members in these discovered paths in order to avoid possible disconnections on the ground when the network is sparsely connected. The different tests of this technique are performed based on NS-2 simulator and the outcomes are compared with those of related on-demand routing protocols dedicated for VANETs. Interesting results are distinguished showing a reduced end-to-end delay and a high delivery ratio, which proving that this heterogeneous communication between vehicles and UAVs is able to extend the network connectivity.
\end{abstract}
\IEEEpeerreviewmaketitle

\section{Introduction}
VANET has seen a growing interest from both academia and industry. Such network can provide a lot of opportunities to design many applications supporting a comfortable and safe driving experience. Additionally, by exchanging messages, this kind of networks offers useful services and entertainment applications to drivers and passengers helping to avoid accidents, traffic jam, and to enhance road capacity. Furthermore, these communications can also include helpful infotainment like the weather, restaurant locations, gas station, parking places, \textit{etc.} A reliable data delivery is considered as a keystone to deploy these aforementioned applications.

Data packet routing plays a basic role to support the performance success of VANETs. Numerous routing challenges need to be addressed in order to adapt the proposed solutions to the unique characteristics of VANETs, especially the movements of vehicles (various speeds and directions). The majority of the proposed reactive routing protocols dedicated for VANETs only discover the existence of routing paths between a pair of vehicles and they are based on a recovery strategy when a link-breakage occurs. As a result, when there is a path failure, a significant delay is diagnosed in the initializing of a new path. Moreover, the majority of these protocols do not take into account whether the discovered paths are dense with vehicles or not in order to increase the probability of a successful data delivery between a pair of source and destination nodes. In other words, when the network on the ground is poorly dense (\textit{i.e.,} sparsely connected), these protocols cannot forward the data packets because of their inefficient recovery strategies (new route discovery) to find a new path, and consequently, the data packets cannot reach their final destinations.

In this paper, we propose a new on-demand routing protocol based on the idea of using UAVs as full members belonging to existing VANETs on the ground and assisting vehicles during the routing process. This proposal is designed for urban cities and is essentially based on the densest (connected), stable, and the fastest paths in terms of delay for the data delivery. A set of techniques is adopted during the discovery process, which aims to reduce the delay of delivery and minimize the routing overhead. A scoring technique based on several criteria is used to select the most appropriate path for the data delivery. Moreover, a maintenance process is deployed in the case of path failures according to the situation of the network as follows: (i) An alternative path can be found at each path failure without re-initiating the discovery process. When the network on the ground is highly fragmented, (ii) UAVs are used to bridge the communication gaps, thus creating new alternative paths.

The remainder of this paper is organized as follows. First of all, we present an overview of relevant related works proposed in the literature in Section \ref{sect2}. In Section \ref{sect3}, the proposed routing protocol is described in details. The performance evaluation and the outcomes of our proposal are presented in Section \ref{sect4}. Finally, Section \ref{sect5} concludes the paper and summarizes some future perspectives.

\section{Related Work}
\label{sect2}
During the last few years, an important number of reactive routing contributions dedicated for VANETs are proposed in the literature. These protocols are based on the broadcasting and the flooding of the entire network in order to discover always new routes \cite{oubbati2017intelligent,venkatramana2017scgrp,lin2017mozo,nzouonta2009vanet,al2018real,khan2018traffic,yan2011agp,he2017delay,mo2006muru}. However, these protocols fail to deliver the data packets when there are no existing paths between a pair of source and destination. Moreover, most of them do not consider the well-balanced density in the discovered paths which is an important parameter to ensure a reliable transmission of packets.

We cite for example, Road-Based using Vehicular Traffic (RBVT) scheme \cite{nzouonta2009vanet}. It uses two different routing strategies: (i) a proactive routing strategy (RBVT-P) and (ii) a reactive routing strategy (RBVT-R). RBVT-R carries out a discovery process in order to find he position of the destination and the set of traversed intersections. This information is reported back to the source based on the greedy forwarding included in the header of a route reply (RR) packet. In the case when the destination received several route discovery (RD), this means that RBVT-R has to select the routing path with a fewer number of transited junctions (\textit{i.e.,} the shortest distance to the source) among many paths discovered beforehand, and then to send the RR through it. At the reception of the RR packet, the source starts the data transmission through the same path crossed by the RR. As a disadvantage, RBVT-R performs the routing decisions based on the shortest distance without considering the traffic density on the segments which may cause a path failure at any time.

Presented in \cite{mo2006muru}, MURU (Multi-hop Routing protocol for Urban VANET) calculates a metric called Expected Disconnection Degree (EDD) which is a probability that a given path might be disconnected during a given period of time. The lower of EDD, the better is the path. The EDD is estimated by combining the vehicle positions, velocities, and trajectories. Consequently, paths along vehicles moving in similar speeds and directions are more stable and therefore more desirable. After calculating the shortest path to the destination, the source node  initializes the route discovery, at the same time the EDD is calculated permanently at each hop and stored in the route request (RREQ). When the destination receives a certain number of RREQ, it chooses the path with the smallest EDD. Nevertheless, this protocol does not take into consideration the density of vehicles considered as an important factor to measure the connectivity and ensure an efficient data delivery.

The authors of \cite{yan2011agp} used a flooding technique to find both the geographical location of the destination and establishing a
well-connected routing path. Therefore, several RREQs reach the destination indicating several routing paths. Then, the destination calculates a multi-criteria weight for each intercepted path (\textit{i.e.,} a set of intersections). The path with the highest weight is selected to make the data transmission. To outperform the mobility of the destination, the forwarder vehicle can use a mobility prediction technique based on the motion direction and the velocity of the destination included in data packets. Notice that this protocol does not consider the real distribution of vehicles on the selected path which may cause a path failure even if this path comprises an important number of vehicles.

The work proposed in \cite{le2006uav}, is originally dedicated for MANETs and it is assisted by UAVs in the sky. UAVs are used as relays assisting the nodes on the ground during the data delivery and enhancing the connectivity particularly when the network is sparsely connected. this protocol uses  the technique of Carry \& Forward only with the UAVs in order to carry the packets to the destination by flying, so this type of protocol is called Disruption Tolerant Network (DTN) and it is used only in the sky. However, on the ground Ad hoc On Demand Distance Vector (AODV) \cite{leonov2018simulation} is applied. The weakest point of this protocol is that UAVs do not exploit the geographical positions of the nodes during the data delivery.

To address the different drawbacks distinguished in the aforementioned protocols, we propose an on-demand routing protocol in which the routing selection process is based on the real distribution of vehicles on the discovered routing paths. Moreover, UAVs can both assist vehicles during the data delivery in the case when there is a weakest connection on the ground and can provide alternative solutions when the network is severely fragmented.

\section{On-Demand Routing for Urban VANETs using Cooperating UAVs}
\label{sect3}
This section describes in detail the different functionalities of the proposed reactive routing approach. The discovery process is exploited to have an accurate knowledge of the traffic density and its real distribution in each routing path. To select the most connected and the well-regulated path, a multi-criteria score is calculated for each discovered path based on several parameters. 

To cope with the intermittent connectivity caused often by the high mobility of vehicles, we propose the use of an intelligent route maintenance process based on alternative paths discovered beforehand and stored in the header of the data packet. UAVs have a major role to assist vehicles on the ground, both in the discovery process and in the case of disconnections.

\subsection{Assumptions}
\begin{figure}[!ht]
\centering
 \includegraphics[scale=0.25]{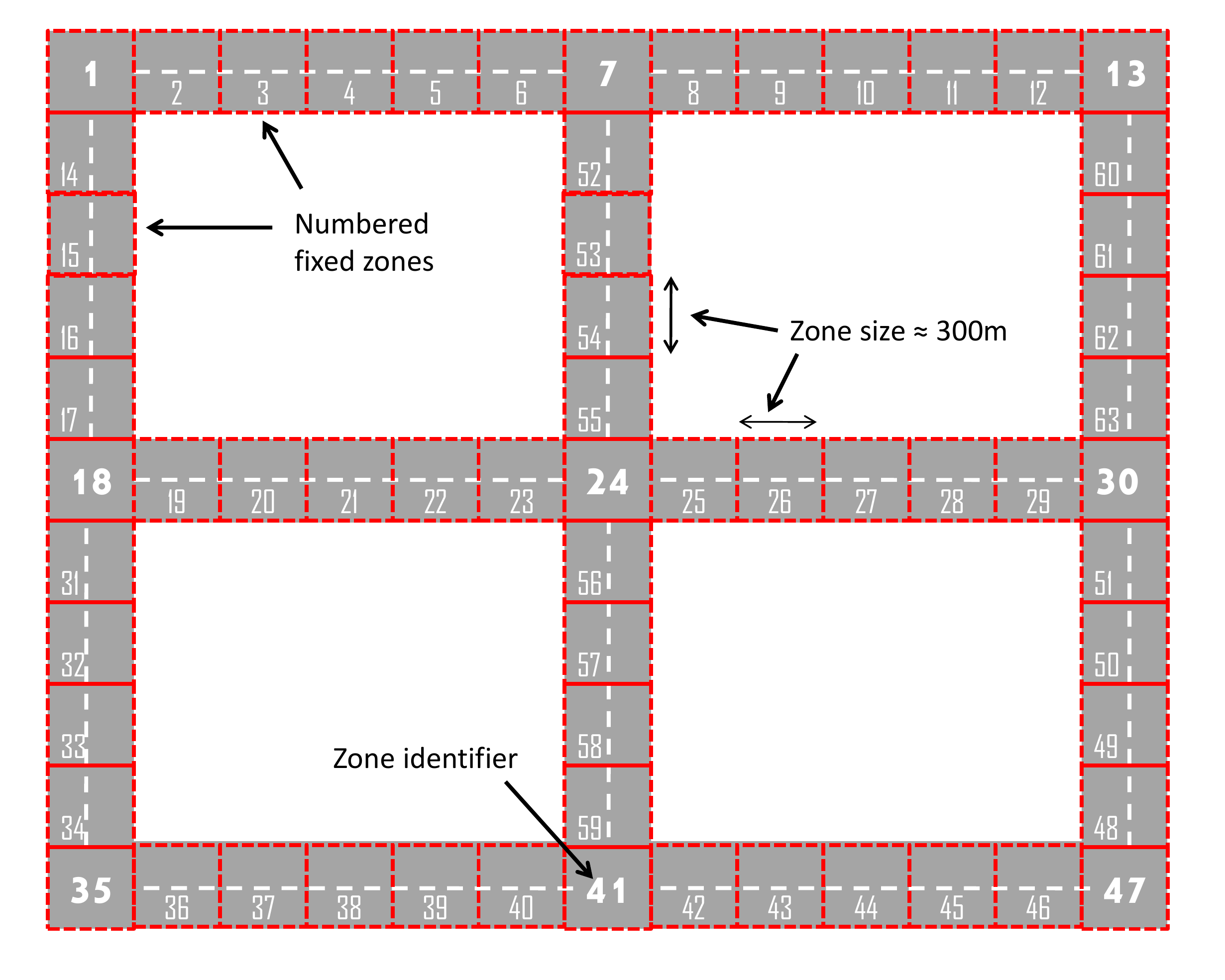}
\caption{Network topology.}
\label{archaisv}
\end{figure}

Our approach consists of vehicles and UAVs equipped with GPS and map. Each node maintains and updates a table of neighbors. We assume that there is no energy constraint for both vehicles and UAVs because they can recharge their batteries from their energy resources (\textit{e.g.,} vehicles’ energy resources or the solar energy). The UAVs are able to communicate with vehicles through wireless interfaces up to a large transmission range with each other, so they will not be affected by obstacles (buildings, \textit{etc.}). In addition, we suppose that the network has a sufficient number of UAVs so that at each moment, at least one UAV hovers an area of four road segments.
Each road segment is split into small fixed zones. The size of each zone depends totally on the communication range of vehicles (Range$\approx$300m). A unique identifier (ID) is given to each zone (\textit{c.f.,} Fig. \ref{archaisv}).

\subsection{Path discovery}
Our protocol is based on the idea of discovering well-connected multiple paths on demand. These paths, which are represented as a sequence of zone ID, are stored in the headers of data packets and are used by the intermediate nodes to send packets geographically between a source and destination nodes. Moreover, they are also used in the path maintenance process.
\subsubsection{The route request (RREQ) packet format}
\begin{figure}[!ht]
 \includegraphics[scale=0.5]{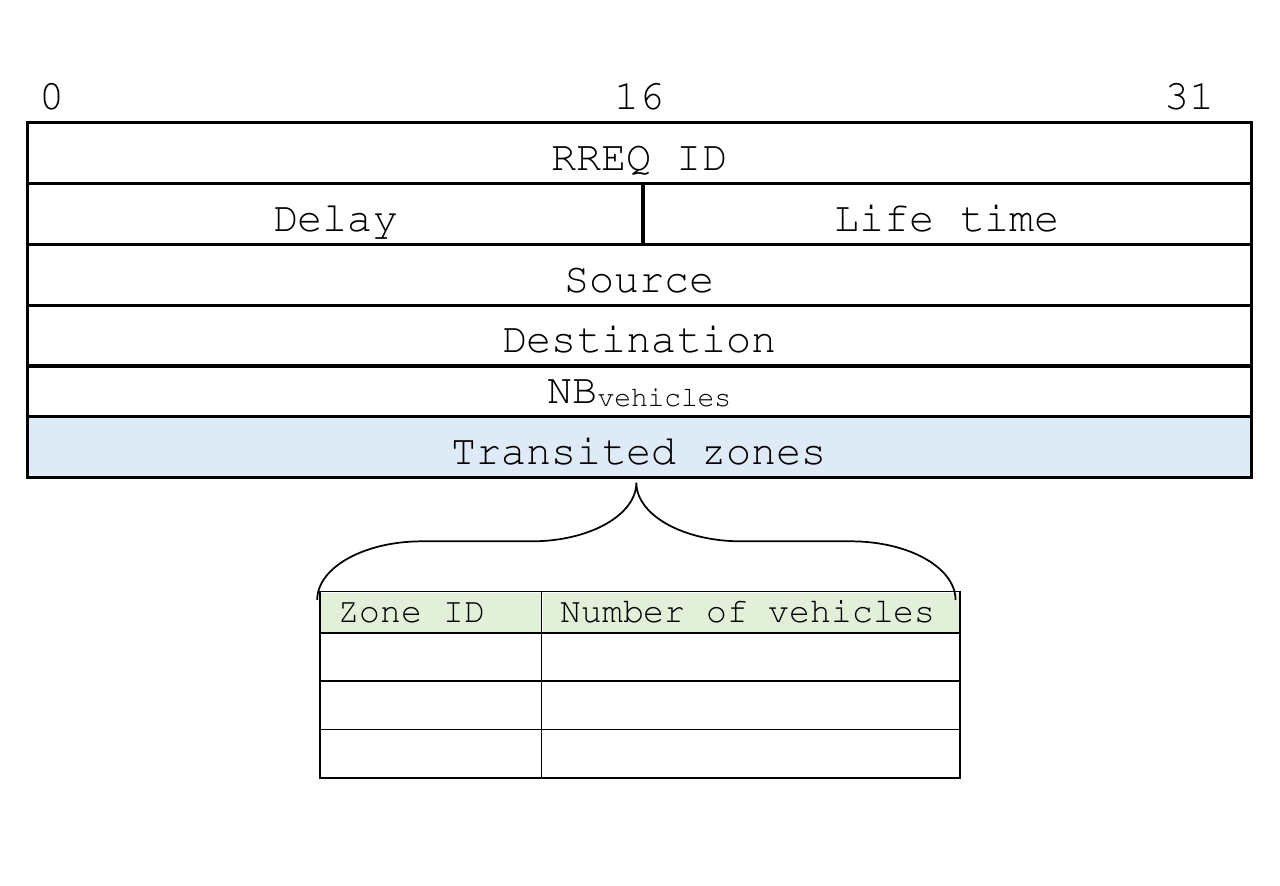}
\caption{The route request (RREQ) format.}
\label{obsss}
\end{figure}

Several fields compose the RREQ packet (\textit{c.f.,} Fig. \ref{obsss}). The $RREQ_{ID}$ field defines the discovery process to which the RREQ packet belongs. The Delay field defines the required time for the data packet to be delivered. $NB_{vehicles}$ field represents the exact number of vehicles that are on the discovered path to the target destination. The $lifetime$ field determines the expiration time of the RREQ packet which is an important parameter limiting the flooding of the entire network. The fields of identifiers of the source vehicle and the target destination. The $Transited~zones$ field is a list of zones until the target destination crossed by the RREQ packet.
\subsubsection{The discovery process}
When a source vehicle wants to send a data packet to a target destination, it initiates a path discovery by flooding a RREQ packet to discover paths toward the destination. The flooding is necessary to get the location of the destination since the proposed protocol does not suppose using a location service. To minimize the impact of the broadcast storm, the $RREQ_{ID}$ field is checked when an intermediate node receives a RREQ packet. If a vehicle finds that the received RREQ has the same $RREQ_{ID}$ with a previously received one, this packet is dropped. Otherwise, the $RREQ_{ID}$ of the received RREQ packet is stored in the $List_{RREQ_{ID}}$ cached in this vehicle or UAV.

The paths are progressively built. Initially, the included path ($Transited~zones$) is an empty list. When the RREQ packet is received by a node (vehicle or UAV) for the first time, it verifies if its zone ID location already exists in the $Transited~zones$ list or not. If so, only the $RREQ_{ID}$ is stored in this vehicle. If not, the $Zone_{ID}$ and the total number of vehicles in its location zone will be added to the enclosed $Transited~zones$ list in the RREQ packet and $NB_vehicles$ will be updated. Then, the vehicle will re-broadcast the RREQ packet to its neighbors.

\begin{figure}[!ht]
\centering
 \includegraphics[scale=0.3]{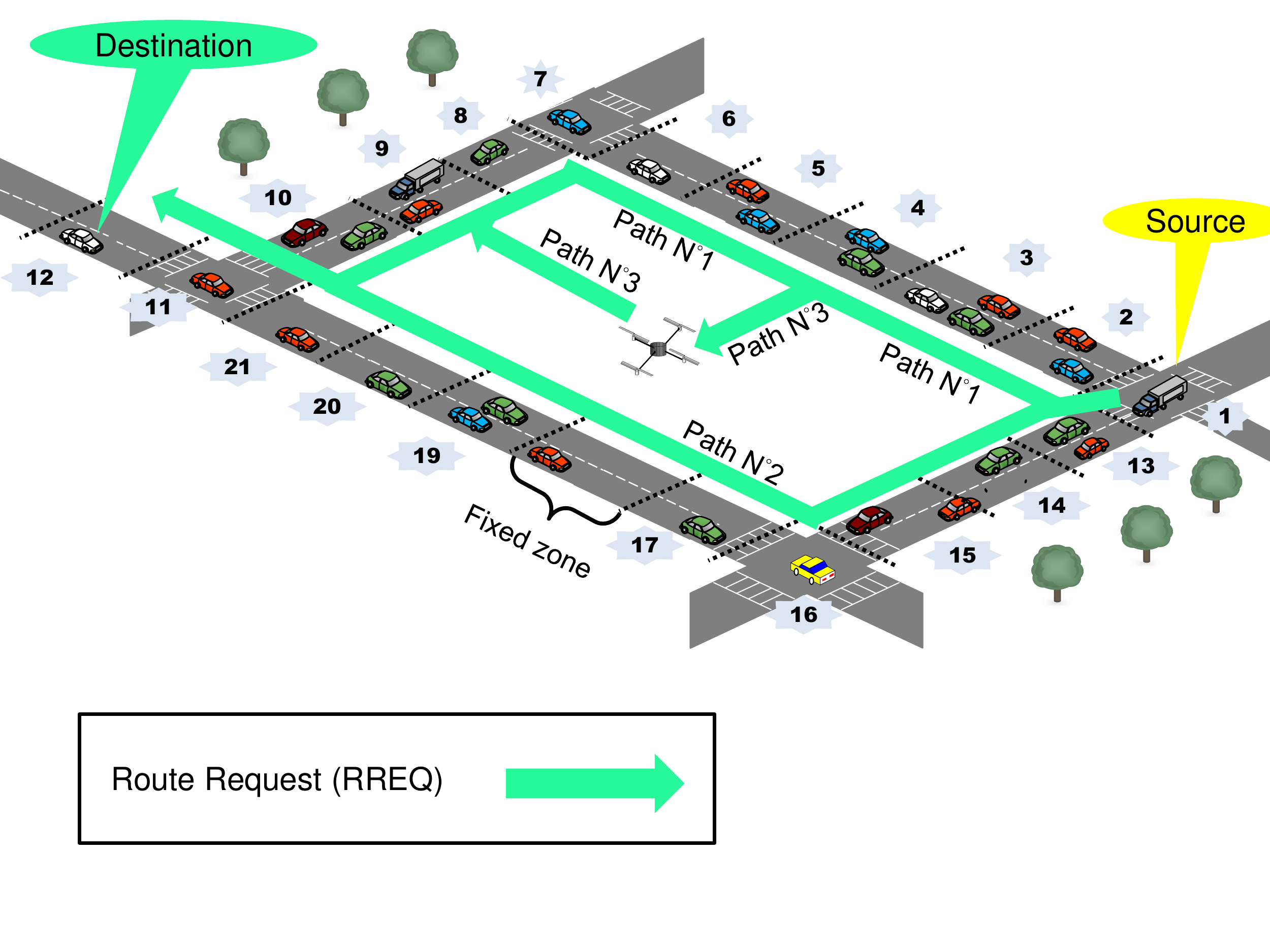}
\caption{Path discovery process.}
\label{disccc}
\end{figure}

We exemplify the discovery process based on Fig. \ref{disccc}. The source vehicle generates a RREQ packet to find paths toward the destination. The source vehicle includes on it the $Zone_{ID}$ where it is located and the number of vehicles that exists within the zone based on its own table of neighbors. Then, the RREQ packet is broadcasted. The same process is carried out by all intermediate nodes, except for UAVs which only add their own IDs in the $Transited~zones$ list.

Finally, when the first RREQ reaches the destination, a timer is then launched to wait a certain time in order to have the knowledge about all the existing paths. The broadcast will be achieved by reaching all the RREQs the destination. The destination has three available paths to the source stored as a list of zones ID in each received RREQ.
\subsection{Path selection}
\begin{figure}[!ht]
\centering
 \includegraphics[scale=0.3]{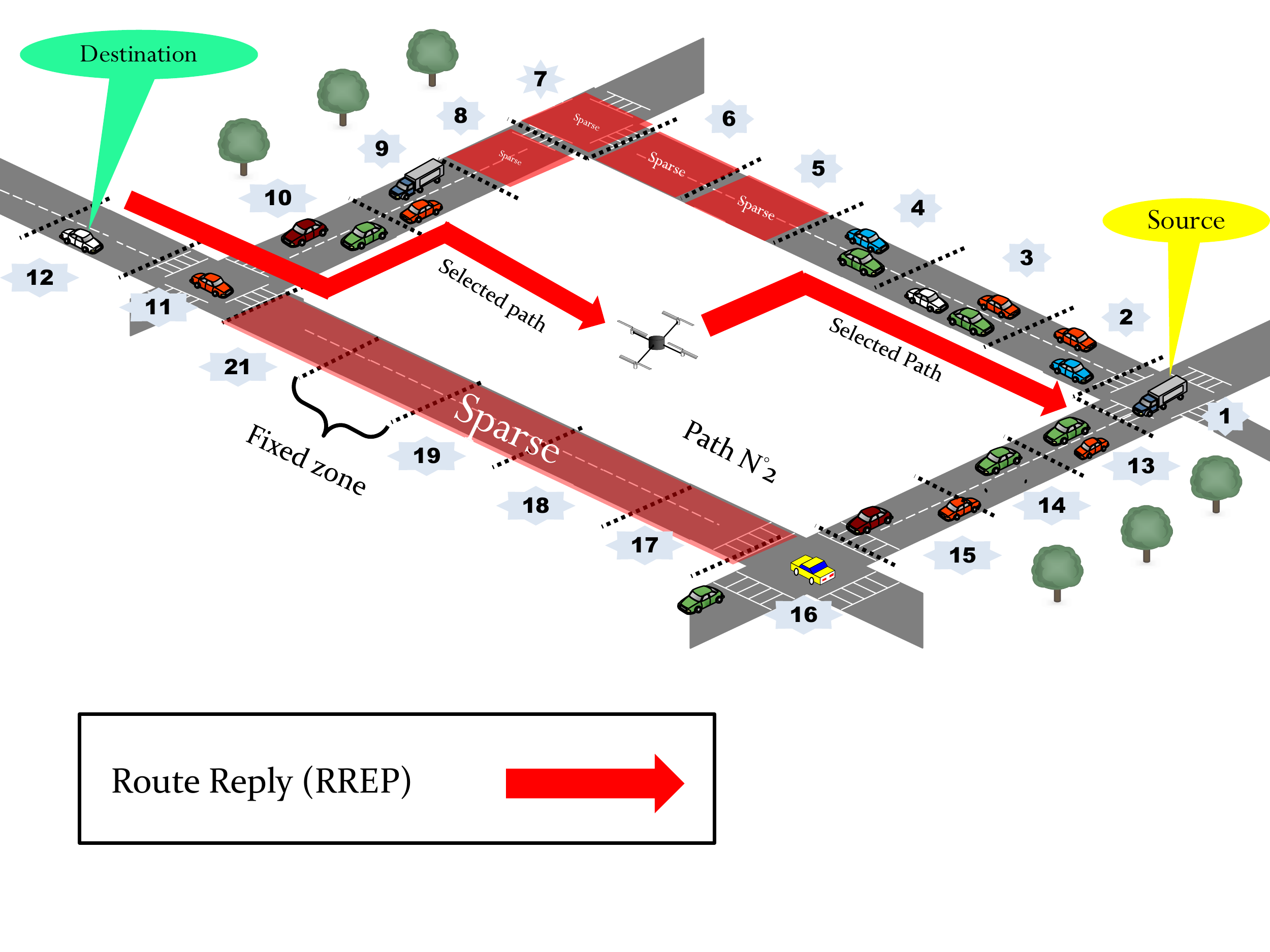}
\caption{Path selection process.}
\label{connectednesssss}
\end{figure}

The selection process is carried out only when the discovery process has finished by discovering at least two routing paths. Indeed, several metrics are calculated for every discovered path based on the received information through the RREQs.  As shown in Fig. \ref{disccc}, the destination has the knowledge of three different paths and accurate parameters about all of them (\textit{see} TABLE \ref{discpaths}). This information is exploited to know the suitable path.
\begin{table}[!ht]
\centering
\caption{Discovered paths.}
\begin{center}
\begin{tabular}{|c|c|c|c|c|c|}
  \hline
  \rowcolor{yellow}
  \multicolumn{2}{|c|}{Path 1} & \multicolumn{2}{|c|}{Path 2} & \multicolumn{2}{|c|}{Path 3}\\
  \hline
  \hline
    \multicolumn{2}{|c|}{$NB_{vehicles}$=19} & \multicolumn{2}{|c|}{$NB_{vehicles}$=15} & \multicolumn{2}{|c|}{$NB_{vehicles}$=15}\\
  \hline
    \multicolumn{2}{|c|}{Delay=1(s)} & \multicolumn{2}{|c|}{Delay=1.5(s)} & \multicolumn{2}{|c|}{Delay=4.5(s)}\\
    \hline
    \rowcolor{pink}
  Zone ID & Density & Zone ID & Density & Zone ID & Density\\
  \hline
  \hline
  1 & 1 & 1 & 1 & 1 & 1\\
  \hline
  2 & 2 & 2 & 2 & 13 & 2\\
  \hline
  3 & 3 & 3 & 3 & 14 & 1\\
  \hline
  4 & 2 & 4 & 2 & 15 & 2\\
  \hline
  5 & 2 & \multicolumn{2}{c|}{\textbf{UAV}} & 16 & 1\\
  \hline
  6 & 1 & 9 & 2 & 17 & 1\\ 
  \hline
  7 & 1 & 10 & 2 & 18 & 1\\
  \hline
  8 & 1 & 11 & 1 & 19 & 2\\
  \hline
  9 & 2 & 12 & 1 & 20 & 1\\
  \hline
  10 & 2 & \multicolumn{2}{c|}{} & 21 & 1\\
  \hline
  11 & 1 & \multicolumn{2}{c|}{} & 11 & 1\\
  \hline
  12 & 1 & \multicolumn{2}{c|}{} & 12 & 1\\ 
  \hline
\end{tabular}
\end{center}
\label{discpaths}
\end{table}\\
Based on the intercepted information shown in TABLE \ref{discpaths}, two main metrics are calculated by the destination per each discovered $Path_i$. First, the average number of vehicles in each path is calculated using the following equation:
\begin{equation}
Average= \frac{1}{N_z} \times NB_{vehicles}
\end{equation}
Where $N_z$ is the total number of traversed zones within a specific path. $NB_{vehicles}$ is the number of vehicles in the $Path_i$. Second, the standard deviation of the zone densities is calculated based on the following equation:
\begin{equation}
S_{deviation}= \sqrt{\left(\frac{1}{N_z}\right)\times\left(\sum_{i=1}^{N_z}{(Z_i-Average)^2}\right)}
\end{equation}
$Z_i$ is the number of vehicles present at a specific $Zone_{ID}$. $S_{deviation}$ shows how the vehicles are balanced in a found path. Usually, a low $S_{deviation}$ indicates that the vehicles are not broadly dispersed around $Average$. However, a high $S_{deviation}$ denotes that the vehicles are more broadly dispersed.\\
A score is calculated for every discovered path by combining the intercepted parameters and the metrics calculated above based on the following equation:
\begin{equation}
Score= \left(\frac{NB_{vehicles}}{Delay}\right) \times \left(\frac{1}{(1+S_{deviation}+HOPs_{UAV})}\right)
\end{equation}
As we can observe, the score has a proportional relationship with $NB_{vehicles}$ within a specific path. However, it has an inverse relationship with $\left(\frac{1}{1+S_{deviation}+HOPs_{UAV}}\right)$ and the $Delay$ where we penalize paths with a large $Delay$ and $S_{deviation}$ and paths which consist of UAVs. Consequently, paths with a low score are undiserable because they can be quickly broken due to the high mobility of vehicles and more particularly UAVs. Once a path that obtains the best score will be selected ($Path_1$ in Fig. \ref{disccc}), a RREP packet will be generated and sent unicastly back to the source using the greedy forwarding technique, \textit{i.e.,} forwarding the RREP to the closest node to the destination along the zones succession of the selected path until it reaches the source (\textit{c.f.,} Fig. \ref{connectednesssss}). It is important to mention that all discovered paths will be copied in the RREP packet in order to be used later in the path maintenance process.

\subsubsection{The route reply (RREP) packet format}
\begin{figure}[!ht]
\centering
 \includegraphics[scale=0.5]{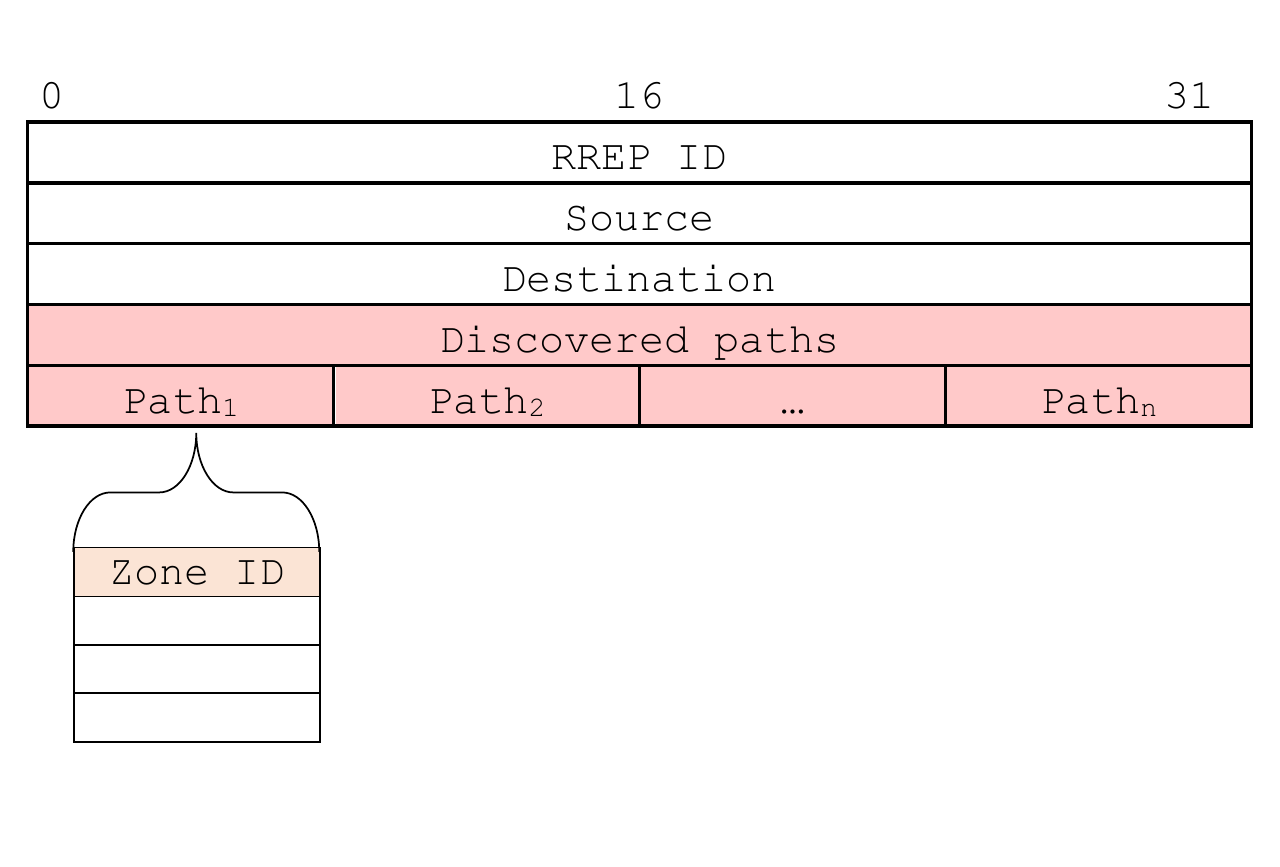}
\caption{The route reply (RREP) format.}
\label{overflsaafown}
\end{figure}

Two information are added by the destination to the RREP packet (\textit{c.f.,} Fig. \ref{overflsaafown}): its geographic location and the discovered paths, which are used further in the path maintenance step.
\subsection{Path maintenance}
Once the source receives the RREP packet, it will copy the discovered paths field into the header of the data packet and starts to send the data packet through the selected path. When the transited path disconnects, the current vehicle or UAV holding the packet has an accurate knowledge of all available paths discovered beforehand which are recorded in the header of the data packet. The adequate path is automatically selected from the alternative paths and the data packet continue its transition. In the case when there are no alternative paths, a route error (RERR) packet including all details about this disconnection is generated and sent back to the source through the initially selected path. At the reception of the RERR packet, the source re-initializes the discovery process.

\begin{figure}[!ht]
\centering
 \includegraphics[scale=0.3]{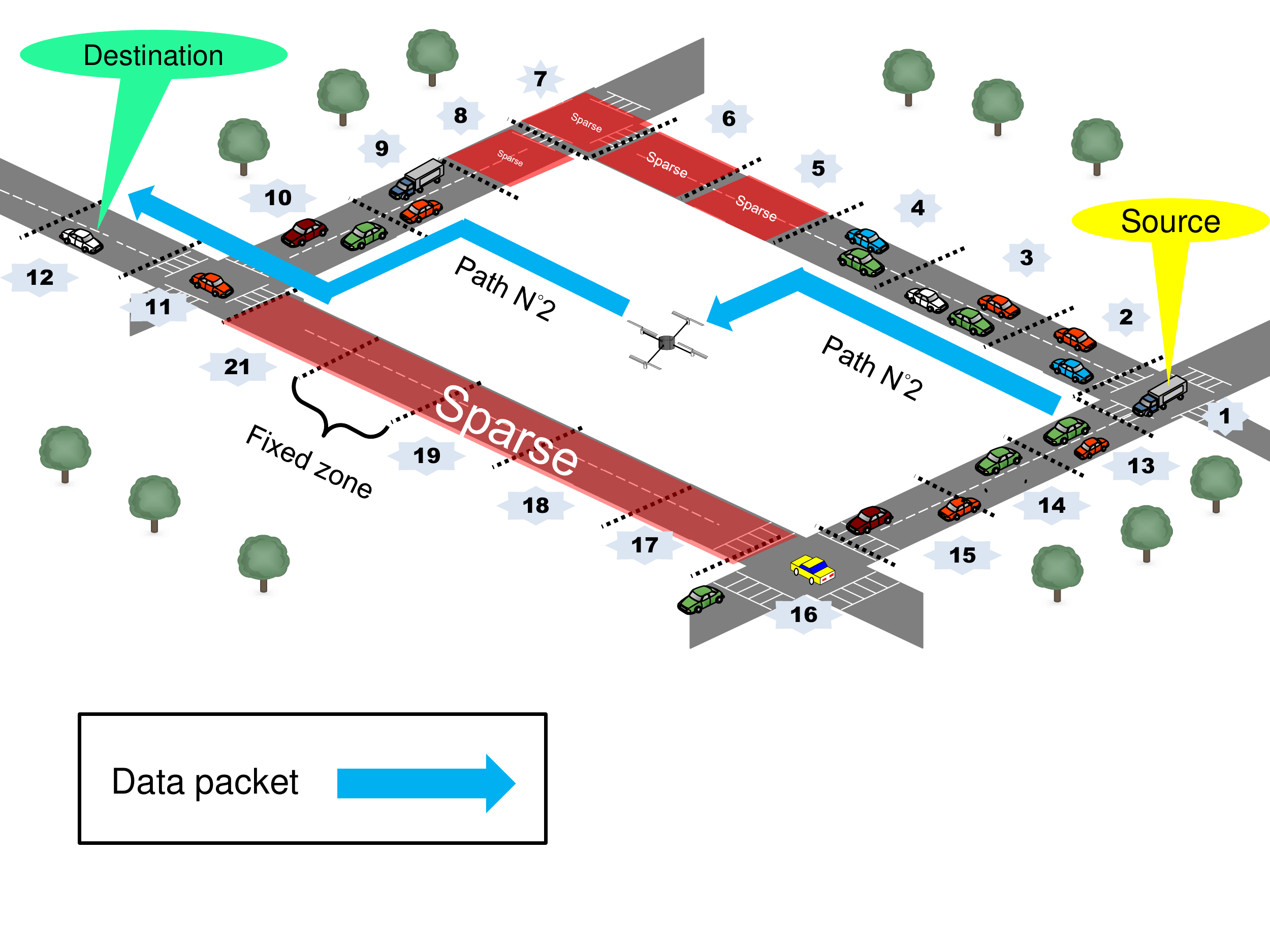}
\caption{Path maintenance process.}
\label{paseffalsse}
\end{figure}

As shown in Fig. \ref{paseffalsse}, the selected $path_1$ is broken at several zones (\textit{i.e.,} 5, 6, 7, and 8) and the current vehicle located at the $Zone_{ID}=4$ detects this disconnection and deducts that the data packet cannot be forwarded through this path. Hence, an alternative path is selected from the available paths stored the header of data packet. After this selection, the UAV is designated as the next forwarder to transmit the data packet through this new path.

It is worth noting that, the UAVs can play the role of a connecting bridge between disconnected clusters. In a general case, additional paths involving UAVs can be found and used to deliver the data packets.
\section{Performance Evaluations}
\label{sect4}
In this section, we present the evaluation of the proposed approach. NS-2 (Network Simulator 2) is used to perform the simulations in order to conduct a series of experiments. Our approach is evaluated and compared to the state of the art routing protocols RBVT-R \cite{nzouonta2009vanet} and AGP \cite{yan2011agp}. 
\subsection{Simulation setup}
A city map of 4$\times$4 $km^2$ comprising 9 intersections is used to make the experiments. The mobility of vehicles and UAVs is generated with VanetMobiSim mobility generator \cite{haerri2016vanetmobisim} and MobiSim mobility generator \cite{mezghani2016mobi}, respectively. TABLE \ref{parametree} summarizes different parameter settings:

\begin{table}[!ht]
\centering
\caption{Simulation parameters}\label{tabparm}
\begin{tabular}{l|r}
  \hline
  \rowcolor{Gray}
  \textbf{Parameter} & \textbf{Value} \\
\hline
  Simulation area & 4000m $\times$ 4000m \\
  \hline
  Number of UAVs & 16 \\
  \hline
  Communication range (vehicles/UAVs) & $\approx$300m \\
  \hline
  MAC Protocol    & 802.11p \\
  \hline
  Frequency Band  & 5.9 GHz \\
  \hline
  Number of packets senders & 35 \\
  \hline
  Data packet size & 1 KB \\
  \hline
  Number of vehicles & 80-200 \\
  \hline
  Vehicle speed & 0-60 km/h \\
  \hline
  UAV speed & 50-120 km/h \\
\hline
  Simulation repeat times & 15 times/scenario \\
\hline
\end{tabular}
\label{parametree}
\end{table}
\begin{figure*}[!ht]
   \centering
    \begin{subfigure}[b]{0.23\textwidth}
        \centering
        \includegraphics[scale=0.35]{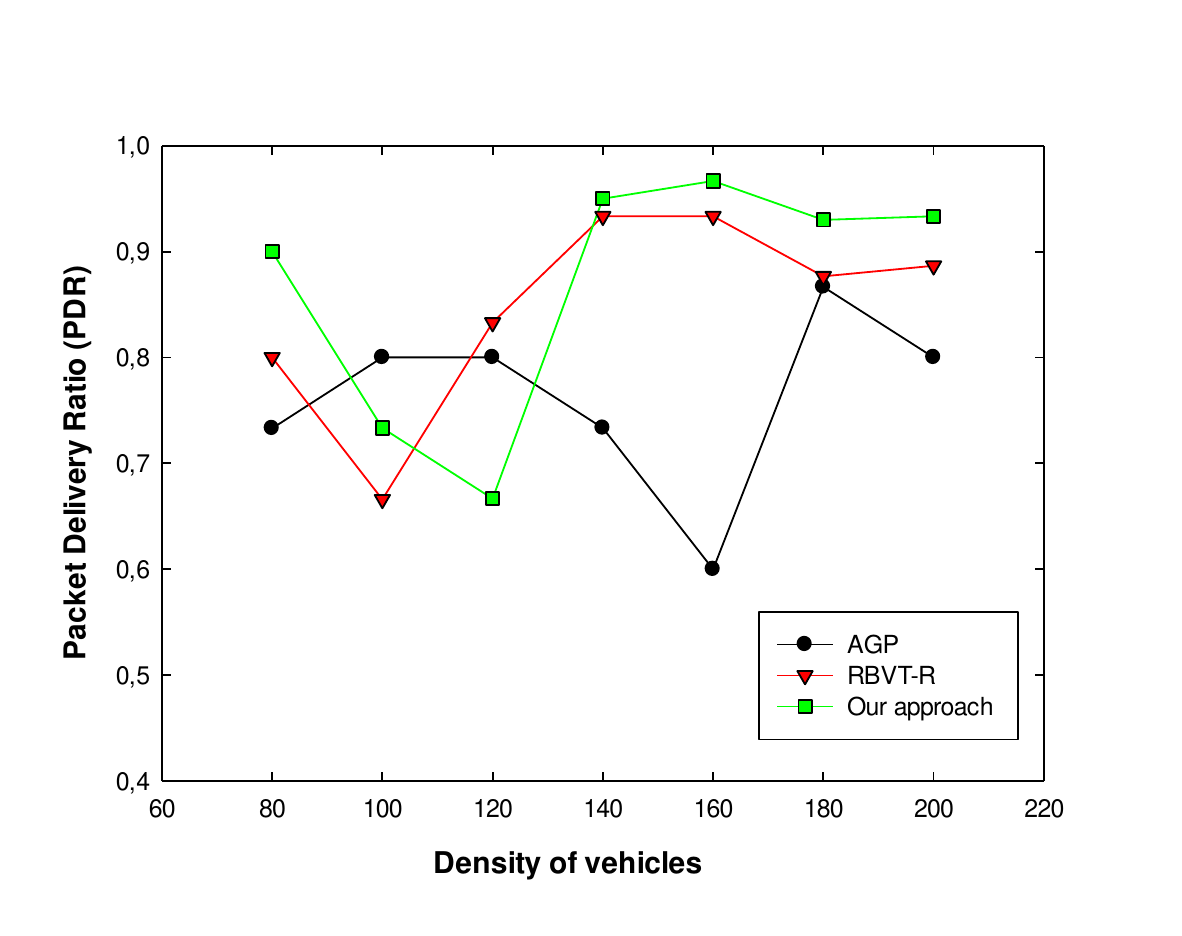}
       \caption{PDR vs. Density}
       \label{pdr}
  \end{subfigure}
 \begin{subfigure}[b]{0.23\textwidth}
      \centering
     \includegraphics[scale=0.35]{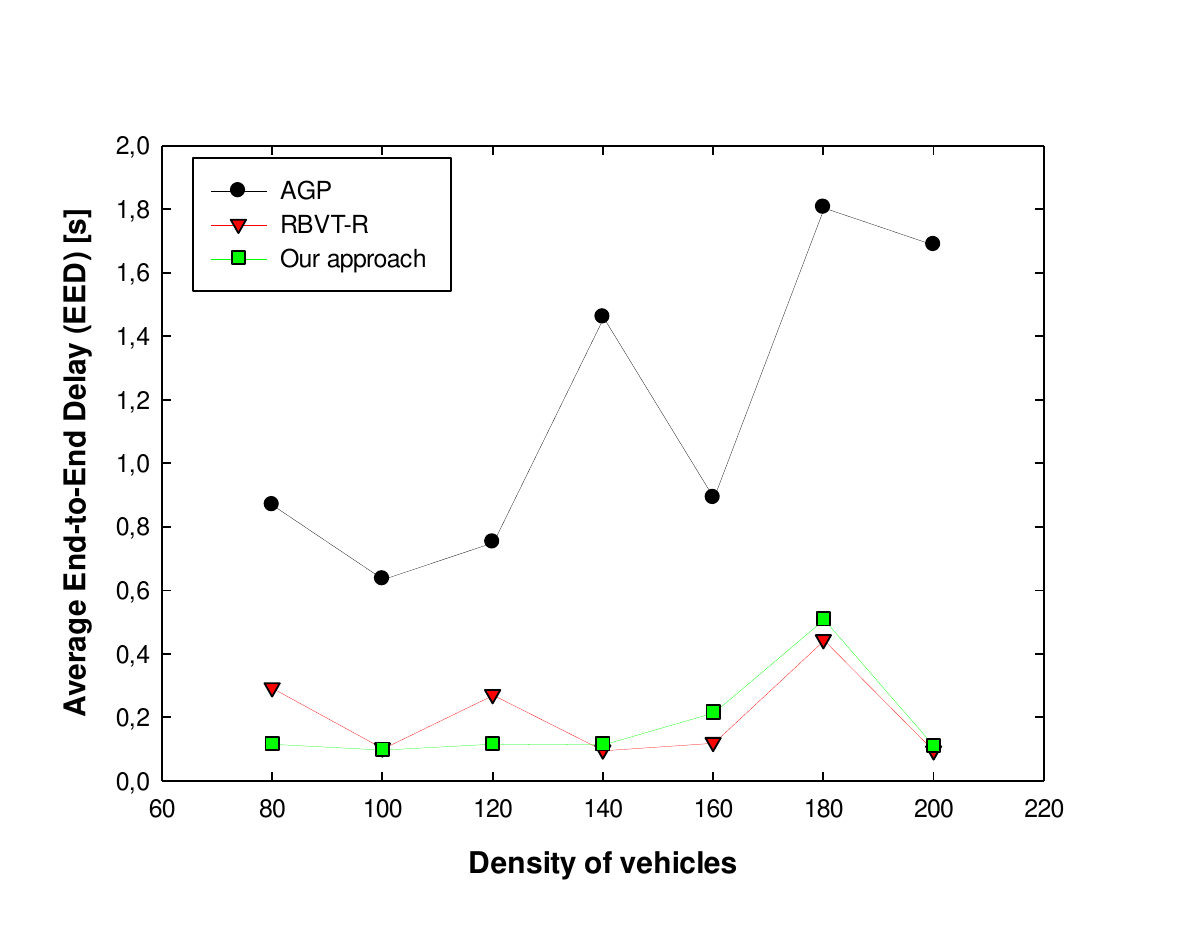}
     \caption{EED vs. Density}
     \label{eed}
  \end{subfigure}
   \begin{subfigure}[b]{0.23\textwidth}
      \centering
       \includegraphics[scale=0.35]{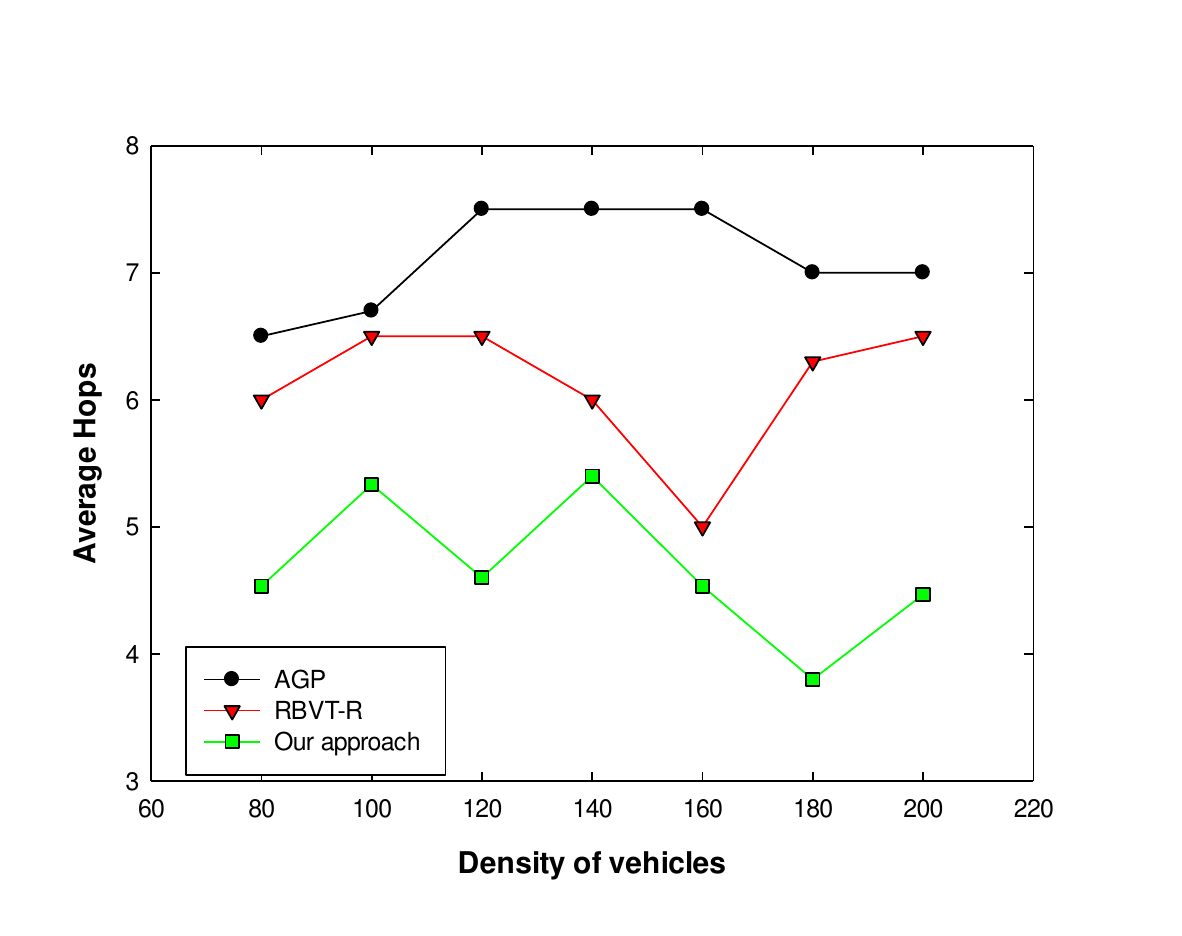}
       \caption{Avg(Hops) vs. Density}
      \label{hops}
  \end{subfigure}
   \begin{subfigure}[b]{0.23\textwidth}
      \centering
       \includegraphics[scale=0.35]{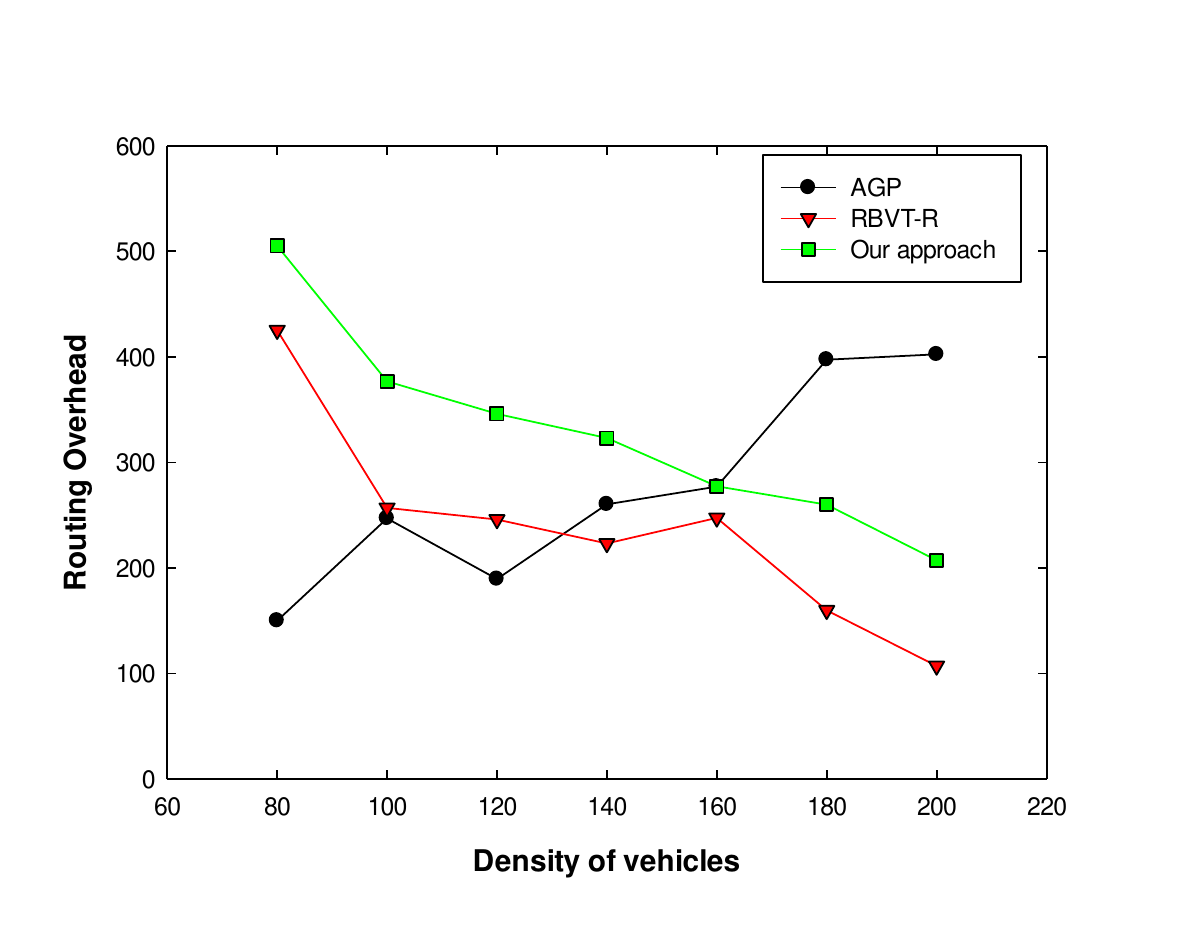}
       \caption{Overhead vs. Density}
      \label{overheadd}
  \end{subfigure}
     \caption{Simulation results}
\end{figure*}

\subsection{Evaluation metric}
Four performance metrics are considered in this experiments:
\begin{enumerate}
\item \textbf{Packet Delivery Ratio (PDR):} is the ratio of delivered data packets at destinations to the total number of packets sent by sources.
\item \textbf{Average Delay (EED):} is the average time taken by a successfully delivered data packets.
\item \textbf{Average number of hops:} is the number of a successfully delivered data packets divided by the total number of hops.
\item \textbf{Routing overhead:} is the number of extra routing packets divided by the successfully delivered data packets at the destination.
\end{enumerate}
\subsection{Results Analysis}
As presented in Fig. (\ref{pdr}), our proposal reached a good results. In fact, we distinguish that the PDR has a proportional relationship with the density of vehicles. This is due to the accurate routing decisions which are based on the traffic density and the real distribution of nodes in each path, which allows to forward data packets efficiently with a minimum packet losses increasing the average delivery ratio. Moreover, in the case of disconnections, we clearly distinguish  the role of UAVs to enhance the connectivity in the network contributing also to this result. RBVT-R performs well than AGP in PDR. As RBVT-R uses a reliable recovery strategy based on a dynamic route updating technique when paths break due to the high mobility of vehicles which is not the case of AGP. RBVT-R demonstrates certain advantages in term of PDR than AGP.

It is clearly seen from Fig. (\ref{eed}) that our approach achieves the lowest delay compared with the other protocols, especially when the network is poorly dense (\textit{i.e.,} Fewer than 140 vehicles). Furthermore, we notice that UAVs significantly reduces the distance transited by the data packet towards the target destination, resulting in a small delay comparing with the other protocols. However, in the high densities (\textit{i.e.,} more than 140 vehicles), RBVT-R outperforms our approach and AGP which results in the routes which remain active for longer periods of time thanks to the reliable route maintenance used by the source vehicle. Unlike RBVT-R, the routes composed of UAVs in our approach are constantly not stable due to high mobility thus causing the triggering of the path discovery process in each path failure. AGP achieves the high delay because it needs more time to discover routing paths and does not have a recovery strategy. 

The average number of hops is illustrated in Fig. (\ref{hops}). We can clearly notice that data packets in our approach require a reduced number of hops to reach their target destinations compared with RBVT-R and AGP. This is because the routing paths in our approach are often composed of UAVs, thus minimizing the transited distance, and consequently, reducing the average number of hops. Furthermore, our approach uses the greedy forwarding technique, which also reduces significantly the number of hops. However, RBVT-R achieves better average of hops compared with AGP, this is because data packets are geographically forwarded along the paths that have a smaller number of intersections towards target destinations which is not the case of AGP. AGP selects paths with a high density of vehicles independently of the number of intersections.

As shown in Fig. (\ref{overheadd}), RBVT-R generates less overhead packets in high density because it does not generate frequent route error (RERR) packets since discovered paths have long lifetime. In low density, AGP performs better than RBVT-R and our approach, because it uses a mobility prediction technique to deal with the frequent topology changes of the network. In addition, AGP does not use RERR packets when there are disconnections in paths and using maps and the real time traffic of vehicles which lead to lower overhead. However, for high density, AGP generates high number of overhead packets caused by the Hello packets. Our approach generates more overhead packets in low densities but reduced progressively with the increase in density of vehicles. This is because our approach generates more RERR packets in low densities because there are no alternative paths. As the number of vehicles increases, our approach always finds alternative paths to recover the selected paths decreasing the number of packets overhead.
\section{Conclusion}
\label{sect5}
This paper has introduced a reactive routing protocol dedicated for urban VANETs where the stability, the connectivity, and the real distribution of vehicles in the path are all considered during the routing decision. The UAVs are involved both in the data delivery and the maintenance process to provide a better connectivity when the network is sparsely connected and to maintain the discovered routing paths when the link failures frequently occur. In addition, whenever there are disconnections or disturbing obstructions, we have distinguished that UAVs can constitute the appropriate assistance in the sky. The simulation outcomes demonstrate that our protocol outperforms existing routing in terms of the packet delivery ratio, the end-to-end delay, and the average number of hops especially the high densities. It is believed that our approach should be able to provide good performances in terms of delivery ratio and average delay in both highway and rural environments. As a future work, we plan to make the mobility of UAVs more controllable in order to enhance the cooperation with vehicles on the ground. Moreover, we plan to enhance this routing protocol by dividing it into two heterogeneous routing components. The first one is executed in the sky exclusively with UAVs and the second one is executed on the ground with vehicles. Furthermore, our recovery strategy will be improved to support the high mobility of UAVs.

\bibliography{mybib}
\bibliographystyle{IEEEtran}
\end{document}